\documentclass[a4paper,11pt]{article}
\pdfoutput=1 

\usepackage{jheppub} 

\usepackage[T1]{fontenc} 
\graphicspath{{fig/}}

\usepackage{ulem}

\title{\boldmath Photoproduction of $P$-wave doubly charmed baryon at future $e^+e^-$ collider}

\author[a,b,1]{Xi-Jie Zhan,\note{Corresponding author.}}
\author[a,b]{Xing-Gang Wu,}
\author[a,b]{Xu-Chang Zheng}

\affiliation[a]{Department of Physics, Chongqing University,\\Chongqing 401331, People's Republic of China}
\affiliation[b]{Chongqing Key Laboratory for Strongly Coupled Physics, Chongqing University,\\Chongqing 401331, People's Republic of China}

\emailAdd{zhanxj@cqu.edu.cn}
\emailAdd{wuxg@cqu.edu.cn}
\emailAdd{zhengxc@cqu.edu.cn}

\abstract{The photoproduction of $P$-wave doubly charmed baryon ($\Xi_{cc}$) is investigated in the context of future high-energy and high-luminosity $e^+e^-$ colliders.
The direct photoproduction via the sub-process $\gamma+\gamma \rightarrow \Xi_{cc} +\bar{c}+\bar{c}$ and the resolved channel $\gamma+g \rightarrow \Xi_{cc} +\bar{c}+\bar{c}$ are considered.
Within the framework of non-relativistic QCD, the calculation encompasses four $P$-wave $(cc)$-diquark configurations: $(cc)_{\bar{\textbf{3}}}[{}^1P_1]$,
$(cc)_{\textbf{6}}[{}^3P_0]$,
$(cc)_{\textbf{6}}[{}^3P_1]$ and
$(cc)_{\textbf{6}}[{}^3P_2]$.
The two $S$-wave states, $(cc)_{\bar{\textbf{3}}}[{}^3S_1]$ and $(cc)_{\textbf{6}}[{}^1S_0]$, are also included for comparison.
The cross sections, as well as the differential distributions involving transverse momentum, rapidity, and angular variables, have been computed.
Numerical results reveal that the resolved photoproduction process plays a significant role and can provide dominant contributions.
The photoproduction rate of the $P$-wave $\Xi_{cc}$ is approximately one order of magnitude lower than that of the $S$-wave.}

\begin{document}
\maketitle
\flushbottom

\section{Introduction}
\label{sec:1}

A doubly heavy baryon is a system composed of two heavy quarks and one light quark. It possesses a simple structure akin to that of a heavy quarkonium, which enables rigorous theoretical analysis. As a result, investigating its production properties is believed to aid in understanding and validating the theory of Quantum Chromodynamics (QCD).
In 2002 and 2005, the SELEX Collaboration reported a possible discovery of the $\Xi_{cc}^+$ baryon\cite{Mattson:2002vu,Ocherashvili:2004hi}.
More recently, in 2017, the LHCb Collaboration confirmed the existence of another type of doubly heavy baryon, $\Xi_{cc}^{++}$, through the decay channel $\Xi_{cc}^{++} \rightarrow \Lambda_{c}^{+} K^{-} \pi^{+} \pi^{+}$ ($\Lambda_{c}^{+} \rightarrow p K^{-} \pi^{+}$). The measured properties of mass and lifetime for this $\Xi_{cc}^{++}$ baryon were found to be in excellent agreement with theoretical calculations.
Subsequently, the LHCb Collaboration further confirmed its existence in the decay $\Xi_{cc}^{++} \rightarrow \pi^+ \Xi_{c}^+$ \cite{Aaij:2018gfl, Aaij:2018wzf}.
The genuine discovery of doubly heavy baryons is poised to trigger a new wave of theoretical research.

Due to the inherent nonrelativistic characteristics coupled with the confinements attributed to strong interactions, the production of doubly heavy baryons encompasses intricate nonperturbative effects that elude computation through conventional perturbative QCD methods. The study conducted by Ma et al. \cite{Ma:2003zk} undertook the task of describing this production process through the utilization of the nonrelativistic QCD (NRQCD) factorization framework \cite{Bodwin:1994jh}. This ingenious framework dissects the procedure into two sequential stages: firstly, the perturbative formation of a heavy-quark pair within a distinct quantum state, often referred to as a diquark, followed by its subsequent nonperturbative transformation into a baryon.
By employing an expansion rooted in the heavy quark's diminished velocity ($v_Q$) within the rest frame of the baryon, the study was able to pinpoint two foremost states of $(cc)$-diquarks at the leading order: ${}_{\bar{\textbf{3}}}[{}^3S_1]$ and ${}_{\textbf{6}}[{}^1S_0]$. These diquark states respectively correspond to the ${ }^{3} S_{1}$ and ${ }^{1} S_{0}$ S-wave configurations, while existing in the $\overline{\textbf{3}}$ and $\textbf{6}$ color states. Accompanying these states are the associated long-distance matrix elements (LDMEs), denoted as $h_{\bar{\textbf{3}}}$ and $h_{\textbf{6}}$, which encapsulate the nonperturbative likelihood of their transition into the baryonic state.

Numerous comprehensive theoretical investigations have delved into the realm of producing doubly heavy baryons~\cite{Baranov:1995rc, Berezhnoy:1996an, Jiang:2012jt, Jiang:2013ej, Chen:2014frw, Yang:2014ita, Yang:2014tca, Zheng:2015ixa, Bi:2017nzv, Sun:2020mvl, Chen:2014hqa, Chen:2019ykv, Chen:2018koh, Martynenko:2014ola, Koshkarev:2016acq, Koshkarev:2016rci, Groote:2017szb, Berezhnoy:2018bde, Brodsky:2017ntu, Berezhnoy:2018krl, Wu:2019gta, Qin:2020zlg, Niu:2018ycb, Niu:2019xuq, Zhang:2022jst, Luo:2022jxq, Luo:2022lcj, Ma:2022cgt}. These investigations have encompassed a range of production mechanisms, including direct processes occurring in $pp$, $ep$, $\gamma\gamma$, and $e^+e^-$ collisions, as well as indirect channels involving the decays of Higgs bosons, $W$ and $Z$ bosons, and top quarks.
For the purpose of simulating hadroproduction in $pp$ collisions, a dedicated generator known as GENXICC \cite{Chang:2007pp, Chang:2009va, Wang:2012vj} has been meticulously developed.

Next-generation $e^+e^-$ colliders have been put forth in recent times, among them being the FCC-ee \cite{FCC:2018evy}, the CEPC \cite{CEPCStudyGroup:2018rmc, CEPCStudyGroup:2018ghi}, and the ILC \cite{ILC:2007bjz, Erler:2000jg}. These cutting-edge colliders are designed to operate at high collision energies and are projected to achieve unparalleled luminosities. As a result of their capabilities, these advanced $e^+e^-$ colliders hold immense potential to serve as exceptional platforms for a wide array of research topics.
The $e^+e^-$ collider provides two main pathways for the direct production of the doubly heavy baryon $\Xi_{cc}$: 
production through $e^+e^-$ annihilation and via the photoproduction mechanism.
In this work, $\Xi_{cc}$ denotes the baryon $\Xi_{ccq}$, 
where $q$ corresponds to an up ($u$), down ($d$), or strange ($s$) quark.
Regarding photoproduction, the $\Xi_{cc}$ baryon can be generated through direct photon-photon fusion, such as $\gamma+\gamma \rightarrow \Xi_{cc} +\bar{c}+\bar{c}$. Beyond direct photoproduction, another category of processes, known as resolved photoproduction \cite{Klasen:2001cu}, exists. In these cases, the photon undergoes a process of resolution, leading to its parton's involvement in the subsequent hard processes.
The resolved photoproduction channels share a comparable order of perturbative expansion with the direct approach, underscoring the need for their incorporation in calculations. 
Earlier investigations \cite{Klasen:2001cu, Li:2009zzu, Zhan:2020ugq, Zhan:2021dlu, Zhan:2022nck, Zhan:2022etq} have indicated that the single resolved channel($\gamma+g$) tend to exert a dominant influence on the photoproduction of heavy quarkonium and doubly heavy baryon at $e^+e^-$ colliders.
The contributions stemming from double resolved photoproduction channels are generally negligible and can be safely ignored.

In this work, we offer an analysis of the photoproduction of $P$-wave doubly charmed baryon at future $e^+e^-$ collider.
Based on the NRQCD factorization framework,
we will consider two types of photoproduction processes: $\gamma+\gamma \rightarrow \Xi_{cc} +\bar{c}+\bar{c}$ and $\gamma +g \rightarrow \Xi_{cc} +\bar{c}+\bar{c}$.
For $(cc)$-diquark, the quantum number is $(cc)_{\bar{\textbf{3}}}[{}^3S_1]$, $(cc)_{\textbf{6}}[{}^1S_0]$, $(cc)_{\bar{\textbf{3}}}[{}^1P_1]$ or $(cc)_{\textbf{6}}[{}^3P_J]$
with $J=0,1,2$.
Section~\ref{sec:2} provides the formulation of the calculation, while Section~\ref{sec:3} presents the numerical results and subsequent discussions. Section~\ref{sec:4} gives a brief summary.

\section{Formulation and calculation}
\label{sec:2}

At future $e^+e^-$ colliders, the initial photons possess the capacity to attain both high energy and luminosity via laser back-scattering(LBS). This phenomenon is expounded upon in the spectrum detailed by Ginzburg et al.~\cite{Ginzburg:1981vm},
\begin{equation}
	f_{\gamma/e}(x)=\frac{1}{N}\left[1-x+\frac{1}{1-x}-4 r(1-r)\right],
\end{equation}
where $x=E_{\gamma} / E_{e}$, $r=x /(x_{m}(1-x))$, and the normalization factor is given by:
\begin{equation}
	N=(1-\frac{4}{x_{m}}-\frac{8}{x_{m}^{2}}) \log(1+x_m)+\frac{1}{2}+\frac{8}{x_{m}}-\frac{1}{2 (1+x_m)^{2}}.
\end{equation}
Here, $x_{m}=4 E_{e} E_{l} \cos ^{2} \frac{\theta}{2}$, with $E_e$ and $E_l$ representing the energies of the incident electron and laser beams, respectively, and $\theta$ denoting the angle between them. The range of energy for the laser back-scattering (LBS) photon is constrained by:
\begin{equation}
	0 \leq x \leq \frac{x_{m}}{1+x_{m}},
\end{equation}
with the optimal value of $x_m$ being $4.83$~\cite{Telnov:1989sd}.

Within the framework of non-relativistic QCD (NRQCD), the photoproduction cross-section of $\Xi_{cc}$ at the $e^+e^-$ collider can be represented as follows:
\begin{eqnarray}
	&&\mathrm{d} \sigma(e^{+} e^{-} \rightarrow e^{+} e^{-} \Xi_{c}+\bar{c}+\bar{c})\nonumber\\
	&&= \int \mathrm{d} x_{1} f_{\gamma / e}(x_{1}) \int \mathrm{d} x_{2} f_{\gamma / e}(x_{2})\nonumber\\
	&& \times\sum_{i, j} \int \mathrm{d} x_{i} f_{i / \gamma}(x_{i}) \int \mathrm{d} x_{j} f_{j / \gamma}(x_{j}) \nonumber\\
	&&\times  \sum_{n} \mathrm{~d} \hat{\sigma}(i j \rightarrow (cc)[n]+\bar{c}+\bar{c})\left\langle \mathcal{O}^{\Xi_{cc}}[n]\right\rangle.
\end{eqnarray}
Here, $f_{\gamma/e}(x)$ represents the energy spectrum of the photon.
$f_{i/\gamma}$ ($i=\gamma,g,u,d,s$) corresponds to the Gl\"uck-Reya-Schienbein (GRS) distribution function of parton $i$ within the photon~\cite{Gluck:1999ub}.
$f_{\gamma/\gamma}(x)=\delta(1-x)$ is utilized for the direct photoproduction process.
$\mathrm{~d} \hat{\sigma}(i j \rightarrow (cc)[n]+\bar{c}+\bar{c})$ denotes the differential partonic cross-section, which is evaluated perturbatively.
For the baryons $\Xi_{cc}$,
$(cc)[n]=(cc)_{\bar{\textbf{3}}}[{}^3S_1]$, $(cc)_{\textbf{6}}[{}^1S_0]$, $(cc)_{\bar{\textbf{3}}}[{}^1P_1]$ or $(cc)_{\textbf{6}}[{}^3P_J]$.
$\langle{\cal O}^{\Xi_{cc}}[n]\rangle=h_{\bar{\textbf{3}}(\textbf{6})}$ denotes the long-distance matrix element (LDME).
Typically, people adopt a potential model approach, drawing parallels with the heavy quarkonium scenario, and introduce and correlate a wave function with $h_{\bar{\textbf{3}}}$\cite{Falk:1993gb,Kiselev:1994pu,Bagan:1994dy,Baranov:1995rc,Berezhnoy:1998aa},
\begin{equation}
	h_{\bar{\textbf{3}}}\simeq|\Psi_{cc}(0)|^2 \,\mathrm{or}\, |\Psi^{\prime}_{cc}(0)|^2.
	\label{eq:h3}
\end{equation}

Regarding $h_{\textbf{6}}$, there exists no specific relation, and for the sake of simplicity, it is assumed to be equal to $h_{\bar{\textbf{3}}}$.
This assumption is rooted in NRQCD's power counting relative to $v_c$, where both $h_{\textbf{6}}$ and $h_{\bar{\textbf{3}}}$ are assigned equivalent orders \cite{Ma:2003zk}.
In accordance with NRQCD, the bound state $\Xi_{c c}$ can be expanded into a series of Fock states:
\begin{equation}
	\label{eq:fock}
	\begin{aligned}
		\left|\Xi_{cc}\right\rangle= c_1(v)|(cc) q\rangle+c_2(v)|(cc) q g\rangle 
		 +c_3(v)|(cc) q g g\rangle+\cdots.
	\end{aligned}
\end{equation}
Because a light quark can easily emit gluons, the constituents in Eq.~(\ref{eq:fock}) carry equivalent significance, specifically, $c_1\sim c_2 \sim c_3$. 
Consider a $cc$ pair in the ${}_{\bar{\textbf{3}}}[{}^3S_1]$ state; one of the heavy quarks can emit a gluon without altering the spin of the heavy quark. Subsequently, this emitted gluon can undergo a splitting process, leading to the formation of a light quark-antiquark pair $q\bar{q}$. This allows the heavy $cc$ pair to interact with the light $q$ and form the composite particle $\Xi_{cc}$.
Similarly, for a $cc$ pair in the ${}_{\textbf{6}}[{}^1S_0]$ state, one of the heavy quarks can emit a gluon while preserving the spin of the heavy quark. This emitted gluon then separates into a light $q\bar{q}$ pair, and the light quarks also have the capability to emit gluons. Consequently, this heavy $cc$ pair can capture a light quark and a gluon to assemble into $\Xi_{Q Q}$.
This is the reason for $h_{\textbf{6}}$ and $h_{\bar{\textbf{3}}}$ holding the same order in $v_c$. For the sake of simplicity, we assume $h_{\textbf{6}}=h_{\bar{\textbf{3}}}$ in the subsequent calculations. It is worth noting that the long-distance matrix elements (LDMEs) function as overarching parameters beyond the perturbative components, indicating that the results can be refined with the acquisition of new LDMEs.

\begin{figure}[t]
	\centering
	\includegraphics[width=0.9\textwidth]{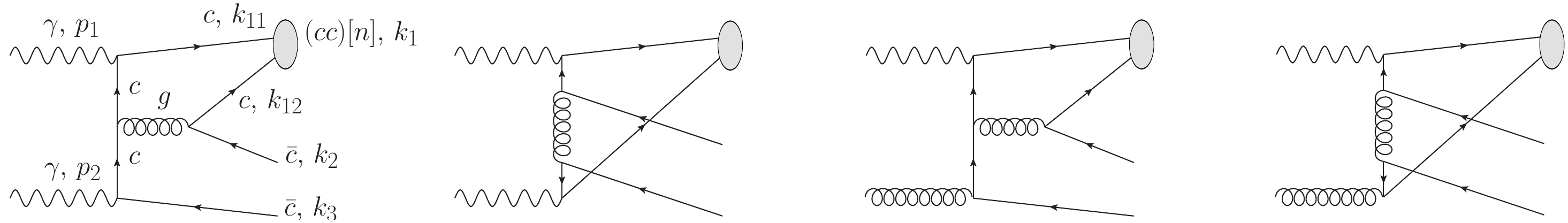}
	\caption{Some Feynman diagrams of the partonic processes for $\Xi_{c c}$ photoproduction. The diagrams are drawn by JaxoDraw~\cite{Binosi:2003yf}.}
	\label{fig:fd}
\end{figure}
For the partonic processes in leading order of $\alpha_s$, there are 40 Feynman diagrams for $\gamma+\gamma \rightarrow \Xi_{c c}+\bar{c}+\bar{c}$ and 48 diagrams for $\gamma+g \rightarrow \Xi_{c c}+\bar{c}+\bar{c}$, as exemplified in Fig.~\ref{fig:fd}.
In practical calculations, we apply charge conjugation $C=-i \gamma^2 \gamma^0$ to reverse the one of the $c\sim\bar{c}$ fermion chains, such as $L_1=\bar{u}_{s_1}(k_{12}) \Gamma_{i+1} S_F(q_i, m_i) \cdots S_F(q_1, m_1) \Gamma_1v_{s_2}(k_2)$.
Here, $\Gamma_i$ represents the interaction vertex, $S_F(q_i, m_i)$ denotes the fermion propagator, where $q_i$ and $m_i$ are the respective momentum and mass parameters. The subscripts $s_1$ and $s_2$ are used for spin indices, while the index $i$ enumerates the fermion propagators $(i=0,1, \ldots)$ along this fermion line.
The conversion obeys:
\begin{equation}
	\begin{aligned}
		v_{s_2}^T(k_2) C & =-\bar{u}_{s_2}(k_2), \\
		C^{-} \bar{u}_{s_1}(k_{12})^T & =v_{s_1}(k_{12}), \\
		C^{-} S_F^T(-q_i, m_i) C & =S_F(q_i, m_i), \\
		C^{-} \Gamma_i^T C & =-\Gamma_i .
	\end{aligned}
\end{equation}
$L_1$ is reversed to:
\begin{equation}
	\begin{aligned}
		L_1 & =L_1^T=v_{s_2}^T(k_2) \Gamma_1^T F_F^T(q_1, m_1) \cdots S_F^T(q_i, m_i) \Gamma_{i+1}^T \bar{u}_{s_1}^T(k_{12}) \\
		& =v_{s_2}^T(k_2) C C^{-} \Gamma_1^T C C^{-} S_F^T(q_1, m_1) C C^{-} \cdots C C^{-} S_F^T(q_i, m_i) C C^{-} \Gamma_{i+1}^T C C^{-} \bar{u}_{s_1}^T(k_{12}) \\
		& =(-1)^{(n+1)} \bar{u}_{s_2}(k_2) \Gamma_1 S_F(-q_1, m_1) \cdots S_F(-q_i, m_i) \Gamma_{i+1} v_{s_1}(k_{12}) ,
	\end{aligned}
\end{equation}
here $n$ is the number of vector vertices in the fermion chain.
Let us take the first diagram in Fig.~\ref{fig:fd} as an example, its amplitudes read:
\begin{equation}
	\begin{aligned}
	M_{1} \sim \,\, & \frac{1}{(k_{12}+k_2)^2}\bar{u}_{s12}(k_{12})\gamma^\mu v_{s 2}(k_2)
	\bar{u}_{s11}(k_{11}) \notin(p_1) \\
	&\frac{\not p_2-\not k_{12}-\not k_2 -\not k_3+m_c}{(\not p_{2}-\not k_{12}-\not k_2-\not k_3)^2-m^2_c}\gamma_\mu \frac{\not p_2-\not k_{3}+m_c}{(\not p_{2}-\not k_{3})^2-m^2_c}\notin(p_2) v_{s 3}(k_3),
	\end{aligned}
\end{equation}
Where $k_{11}$ and $k_{12}$ are momenta of the two c quarks.
In the diquark state, their relative momentum $q$ is small and we can set $k_{11}=\frac{m_c}{M_{cc}}k_1+q$ and $k_{12}=\frac{m_c}{M_{cc}}k_1-q$
with $k_1$ being the momentum of the diquark. 
Here the diquark mass $M_{cc}=2m_c$ is adopted in order to ensure the gauge invariance of the amplitude.
After reverse the first fermion line, the amplitude becomes:
\begin{equation}
	\begin{aligned}
		M_{1} \sim \,\, & -\frac{1}{(k_{12}+k_2)^2}\bar{u}_{s2}(k_{2})\gamma^\mu v_{s 12}(k_{12})
		\bar{u}_{s11}(k_{11}) \notin(p_1) \\
		&\frac{\not p_2-\not k_{12}-\not k_2 -\not k_3+m_c}{(\not p_{2}-\not k_{12}-\not k_2-\not k_3)^2-m^2_c}\gamma_\mu \frac{\not p_2-\not k_{3}+m_c}{(\not p_{2}-\not k_{3})^2-m^2_c}\notin(p_2) v_{s 3}(k_3).
	\end{aligned}
\end{equation}
Now we can replace $v_{s {12}}(k_{12}) \bar{u}_{s {11}}(k_{11})$ by the spin projector for $(c\bar{c})[n]$ to get the amplitude of $\gamma + \gamma \rightarrow (cc)[n] + \bar{c} +\bar{c}$.
The spin projector takes the form of 
\begin{equation}
	\Pi_{k_1}(q)=\frac{-\sqrt{M_{c c}}}{4 m^2_c}(\not k_{12}-m_c) \gamma^5(\not k_{11}+m_c),
\end{equation}
\begin{equation}
	\Pi_{k_1}^\beta(q)=\frac{-\sqrt{M_{c c}}}{4 m^2_c}(\not k_{12}-m_c) \gamma^\beta(\not k_{11}+m_c),
\end{equation}
for $n={}^1S_0$ and ${}^3S_1$ respectively. The amplitudes of the $P$-wave production can be obtained via the derivation of the $S$-wave expression, i.e., 
\begin{equation}
	\begin{aligned}
		M_{1}[{ }^1 P_1]\sim\, & \varepsilon_\alpha^l(k_1) \frac{d}{d q_\alpha}[-\frac{1}{(k_{12}+k_2)^2}\bar{u}_{s2}(k_{2})\gamma^\mu \frac{-\sqrt{M_{c c}}}{4 m^2_c}(\not k_{12}-m_c) \gamma^5(\not k_{11}+m_c)\notin(p_1) \\
		&\frac{\not p_2-\not k_{12}-\not k_2 -\not k_3+m_c}{(\not p_{2}-\not k_{12}-\not k_2-\not k_3)^2-m^2_c}\gamma_\mu \frac{\not p_2-\not k_{3}+m_c}{(\not p_{2}-\not k_{3})^2-m^2_c}\notin(p_2) v_{s 3}(k_3)]|_{q=0},
	\end{aligned}
\end{equation}
\begin{equation}
	\begin{aligned}
		M_{1}[{ }^3 P_J]\sim\, & \varepsilon_{\alpha \beta}^J(k_1) \frac{d}{d q_\alpha}[-\frac{1}{(k_{12}+k_2)^2}\bar{u}_{s2}(k_{2})\gamma^\mu \frac{-\sqrt{M_{c c}}}{4 m^2_c}(\not k_{12}-m_c) \gamma^\beta(\not k_{11}+m_c)\notin(p_1) \\
		&\frac{\not p_2-\not k_{12}-\not k_2 -\not k_3+m_c}{(\not p_{2}-\not k_{12}-\not k_2-\not k_3)^2-m^2_c}\gamma_\mu \frac{\not p_2-\not k_{3}+m_c}{(\not p_{2}-\not k_{3})^2-m^2_c}\notin(p_2) v_{s 3}(k_3)]|_{q=0},
	\end{aligned}
\end{equation}
Here, $\varepsilon_\beta^s(k_1)$ or $\varepsilon_\alpha^l(k_1)$ represents the polarization vector associated with the spin or orbital angular momentum of the diquark in the spin triplet $S$-state or spin singlet $P$-state. $\varepsilon_{\alpha \beta}^J(k_1)$ corresponds to the polarization tensor for the spin triplet $P$-wave states, where $J$ can be 0, 1, or 2. To determine the suitable total angular momentum, we appropriately perform the polarization sum.
The summation over polarization vectors is carried out as follows:
\begin{equation}
	\sum_{r_z} \varepsilon_\alpha^r \varepsilon_{\alpha^{\prime}}^{r *}=\Pi_{\alpha \alpha^{\prime}}
\end{equation}
The summation over polarization tensors is conducted as:
\begin{equation}
	\varepsilon_{\alpha \beta}^0 \varepsilon_{\alpha^{\prime} \beta^{\prime}}^{0 *}=\frac{1}{3} \Pi_{\alpha \beta} \Pi_{\alpha^{\prime} \beta^{\prime}}
\end{equation}
\begin{equation}
	\sum_{J_z} \varepsilon_{\alpha \beta}^1 \varepsilon_{\alpha^{\prime} \beta^{\prime}}^{1 *}=\frac{1}{2}(\Pi_{\alpha \alpha^{\prime}} \Pi_{\beta \beta^{\prime}}-\Pi_{\alpha \beta^{\prime}} \Pi_{\alpha^{\prime} \beta})
\end{equation}
\begin{equation}
	\sum_{J_z} \varepsilon_{\alpha \beta}^2 \varepsilon_{\alpha^{\prime} \beta^{\prime}}^{2 *}=\frac{1}{2}(\Pi_{\alpha \alpha^{\prime}} \Pi_{\beta \beta^{\prime}}+\Pi_{\alpha \beta^{\prime}} \Pi_{\alpha^{\prime} \beta})-\frac{1}{3} \Pi_{\alpha \beta} \Pi_{\alpha^{\prime} \beta^{\prime}} ,
\end{equation}
with the definition
\begin{equation}
	\Pi_{\alpha \beta}=-g_{\alpha \beta}+\frac{k_{1 \alpha} k_{1 \beta}}{M_{c c}^2}.
\end{equation}

For the processes $\gamma + \gamma \rightarrow (cc)[n] +\bar{c}+\bar{c}$, the color factor $\mathcal{C}_{i j, k}$ is universal for all the Feynman diagrams and $\mathcal{C}_{i j, k}=\mathcal{N} \times \sum_{a, m, n}(T^a)_{m i}(T^a)_{n j} \times G_{m n k}$.
Here $\mathcal{N}=1/\sqrt{2}$ is the normalization constant, $i,j,m,n$ are color indices of four heavy quarks and $k$ denotes the color index of the diquark.
$G_{m n k}$ corresponds to either the antisymmetric function $\varepsilon_{m n k}$ for the color anti-triplet state, or the symmetric function $f_{m n k}$ for the color sextuplet state,
and they obey
\begin{equation}
	\varepsilon_{m n k} \varepsilon_{m^{\prime} n^{\prime} k}=\delta_{m m^{\prime}} \delta_{n n^{\prime}}-\delta_{m n^{\prime}} \delta_{n m^{\prime}},
\end{equation}
\begin{equation}
	f_{m n k} f_{m^{\prime} n^{\prime} k}=\delta_{m m^{\prime}} \delta_{n n^{\prime}}+\delta_{m n^{\prime}} \delta_{n m^{\prime}}.
\end{equation}
The color factors of diagrams for $\gamma +g \rightarrow (cc)[n]+\bar{c}+\bar{c}$ are not the same; they need to be calculated individually.

\section{Numerical results and discussions}
\label{sec:3}

In the calculation, we adopt the wave functions at the origin from \cite{Kiselev:2002iy} as $|\Psi(0)|^2 = 0.0218 \, \mathrm{GeV^3}$ and  $|\Psi^{\prime}(0)|^2 = 2.48\times10^{-3} \, \mathrm{GeV^5}$.
The quark mass is set as $m_c = M_{\Xi_{cc}}/2 = 1.8 \, \mathrm{GeV}$.
The fine structure constant is assigned the value $\alpha = 1/137$.
Regarding the strong coupling constant, we utilize the one-loop running formulation. The renormalization scale is typically taken as the transverse mass of $\Xi_{cc}$, specifically $\mu = \sqrt{M^2_{\Xi_{cc}} + p^2_t}$, where $p_t$ denotes the transverse momentum of the particle.

Table~\ref{tab:cross-section1} lists cross sections of different photoproduction processes, where three collision energies, $\sqrt{S}=250,500,1000\mathrm{GeV}$, are adopted.
From the table, it is evident that the contribution from production channel $\gamma+\gamma$ decreases as the collision energy increases, while the contribution from production channel $\gamma+g$ increases with higher energy levels.
As a cumulative result, the total production cross-section increases with the growth of collision energy.
At $\sqrt{S}=250\,\mathrm{GeV}$, the contribution from the $\gamma+\gamma$ channel is comparable to that of the $\gamma+g$ channel. However, as the collision energy increases, the scenario changes. The $\gamma+g$ channel begins to dominate the photoproduction process. 
Specifically, at $\sqrt{S}=500\,\mathrm{GeV}$, the $\gamma+\gamma$ channel provides 20\% of the contribution, while the $\gamma+g$ channel contributes 80\%.
Table~\ref{tab:cross-section1} clearly emphasizes the significance of the single resolved photoproduction channel.
\begin{table}[tp]
	\centering
	\begin{tabular}{|c|ccc|}
		\hline
		$\sqrt{S}(\mathrm{GeV})$ & $\gamma+\gamma \rightarrow \Xi_{c c}+\bar{c}+\bar{c}$ & 
 $\gamma+g \rightarrow \Xi_{c c}+\bar{c}+\bar{c}$ & total\\
		\hline
		250 & $228.61$ & $238.90$ & $467.50$\\
		500 & $101.34$ & $411.50$ & $512.84$\\
		1000 & $40.94$ & $659.81$ & $700.75$\\
		\hline
	\end{tabular}
	\caption{\label{tab:cross-section1}The integrated cross sections (in unit of fb) from different channels for $\Xi_{c c}$ photoproduction under various collision energies at future $e^+e^-$ collider.
	The contributions from the $S$-wave and $P$-wave have been combined.}
\end{table}
\begin{table}
	\centering
	\begin{tabular}{|c|cccccc|}
		\hline
		$\sqrt{S}(\mathrm{GeV})$ & $(cc)_{\bar{\textbf{3}}}[{}^3S_1]$ & $(cc)_{\textbf{6}}[{}^1S_0]$
		& $(cc)_{\bar{\textbf{3}}}[{}^1P_1]$
		& $(cc)_{\textbf{6}}[{}^3P_0]$ & $(cc)_{\textbf{6}}[{}^3P_1]$ & $(cc)_{\textbf{6}}[{}^3P_2]$\\
		\hline
		250 & $407.43$ & $36.32$ & $8.81$ & $1.75$ & $2.44$ & $10.75$\\
		500 & $442.71$ & $41.24$ & $9.42$ & $2.12$ & $3.03$ & $14.32$ \\
		1000 & $603.0$ & $57.05$ & $12.66$ & $2.97$ & $4.33$ & $20.72$ \\
		\hline
	\end{tabular}
	\caption{\label{tab:cross-section2}The integrated cross sections (in unit of fb) of different intermediate diquark states for $\Xi_{c c}$ photoproduction under various collision energies at future $e^+e^-$ collider.}
\end{table}

Similarly, for the three collision energies, cross sections of various intermediate diquark states are provided in Table~\ref{tab:cross-section2}.
These results reveal that the contribution of $S$-wave diquarks is significantly larger than that of the $P$-wave diquarks.
At $\sqrt{S}=250\,\mathrm{GeV}$, the ratio between the contributions from the $S$-wave and $P$-wave is 18.7:1.
At $\sqrt{S}=500\mathrm{\,GeV}$, the ratio becomes 16.8:1, meaning that the $P$-wave contributions constitute 5.6\% of the total production.
Among the $P$-wave states, the contribution of the $(cc)_{\textbf{6}}[{}^3P_2]$ diquark is the largest. Their ratio is as follows: $(cc)_{\bar{\textbf{3}}}[{}^1P_1]~$:$~(cc)_{\textbf{6}}[{}^3P_0]~$:$~(cc)_{\textbf{6}}[{}^3P_1]~$:$~(cc)_{\textbf{6}}[{}^3P_2]$=1:0.23:0.32:1.52 at $\sqrt{S}=500\,\mathrm{GeV}$.
Assuming an integrated luminosity of $\mathcal{O}(10^4)\mathrm{~fb^{-1}}$ at future $e^+e^-$ colliders and summing up the contributions from all $P$-wave excited baryons, approximately $2.9\times10^5$ $P$-wave $\Xi_{cc}$ baryons would be generated, given a collision energy of $\sqrt{S}=500\mathrm{~GeV}$. The $P$-wave doubly charmed baryons are likely to decay to the ground state with almost 100\% probability, making them additional sources of ground-state baryons.

\begin{figure}
	\centering 
	\includegraphics[width=.31\textwidth]{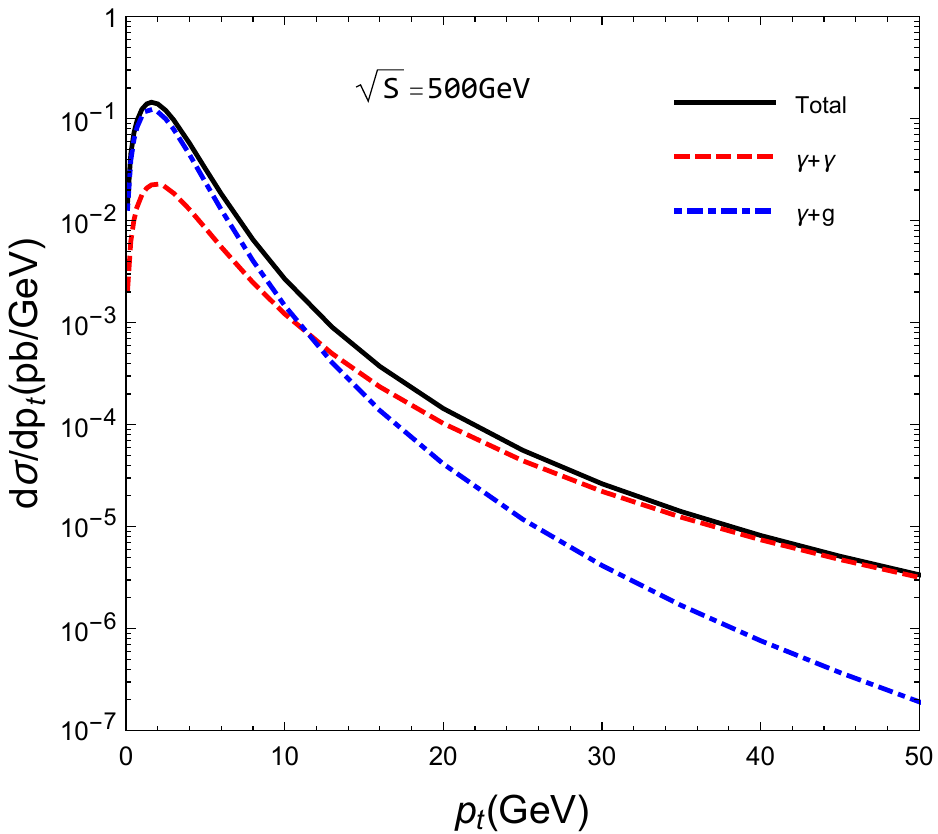}
	\includegraphics[width=.31\textwidth]{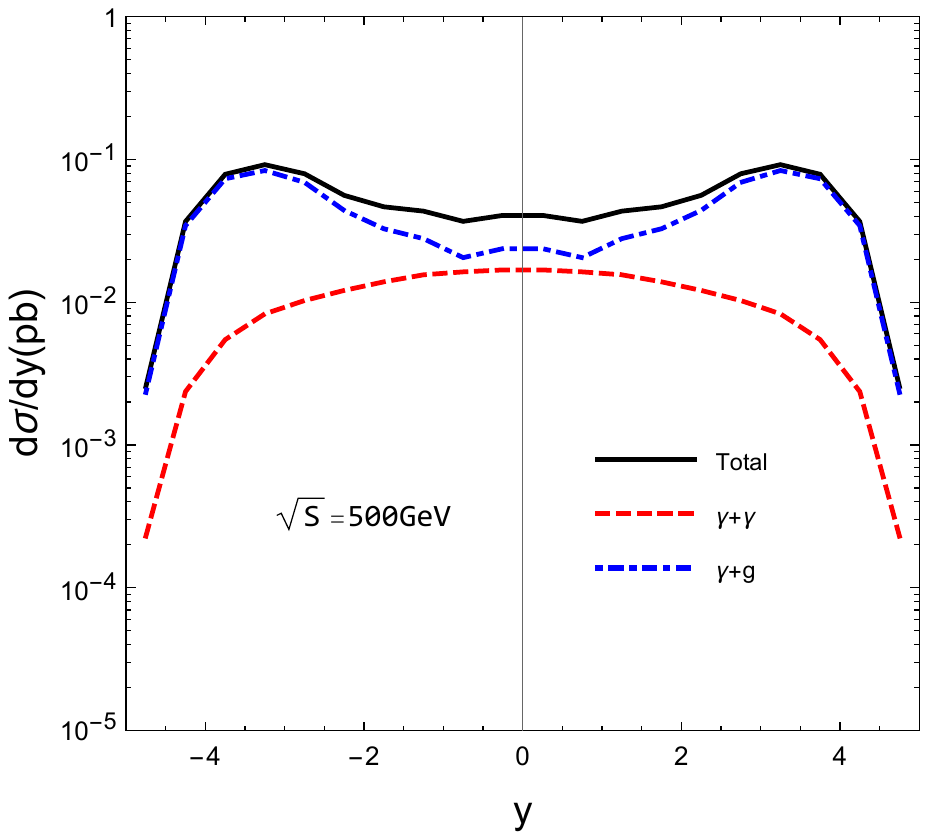}
	\includegraphics[width=.31\textwidth]{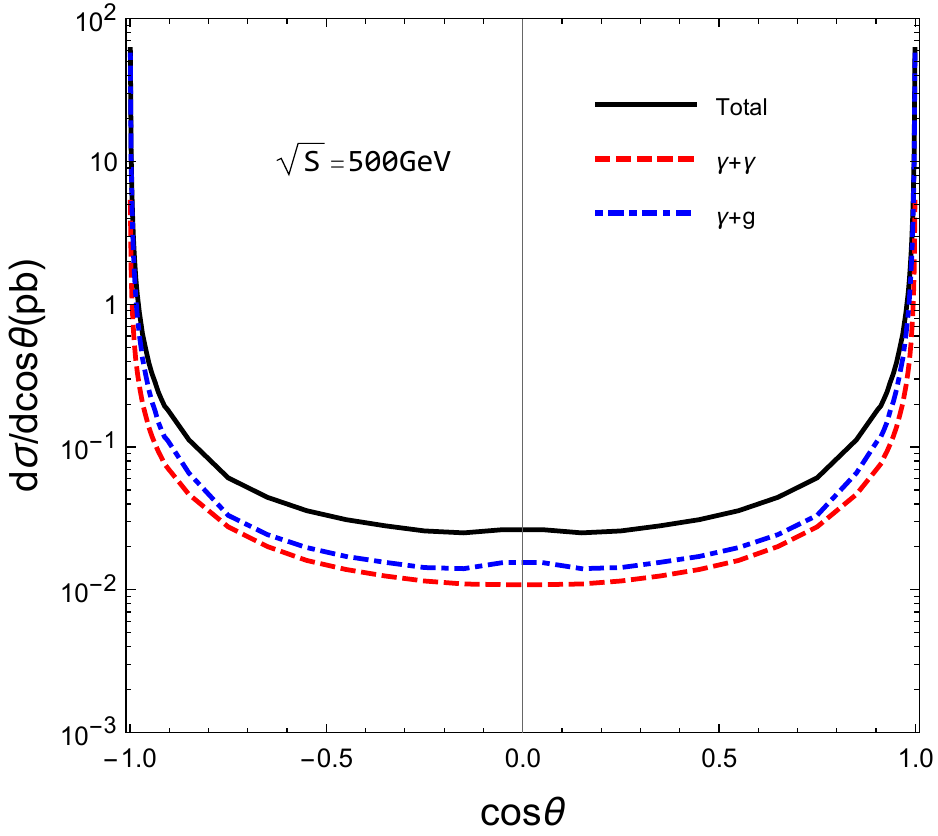}	
	\caption{\label{fig:32} Kinematic distributions for the photoproduction of $\Xi_{cc}$ at future $e^+e^-$ collider($\sqrt{S}=$$500$ $\mathrm{GeV}$). Contributions from different channels are displayed individually.}
\end{figure}
\begin{figure}
	\centering
	\includegraphics[width=.31\textwidth]{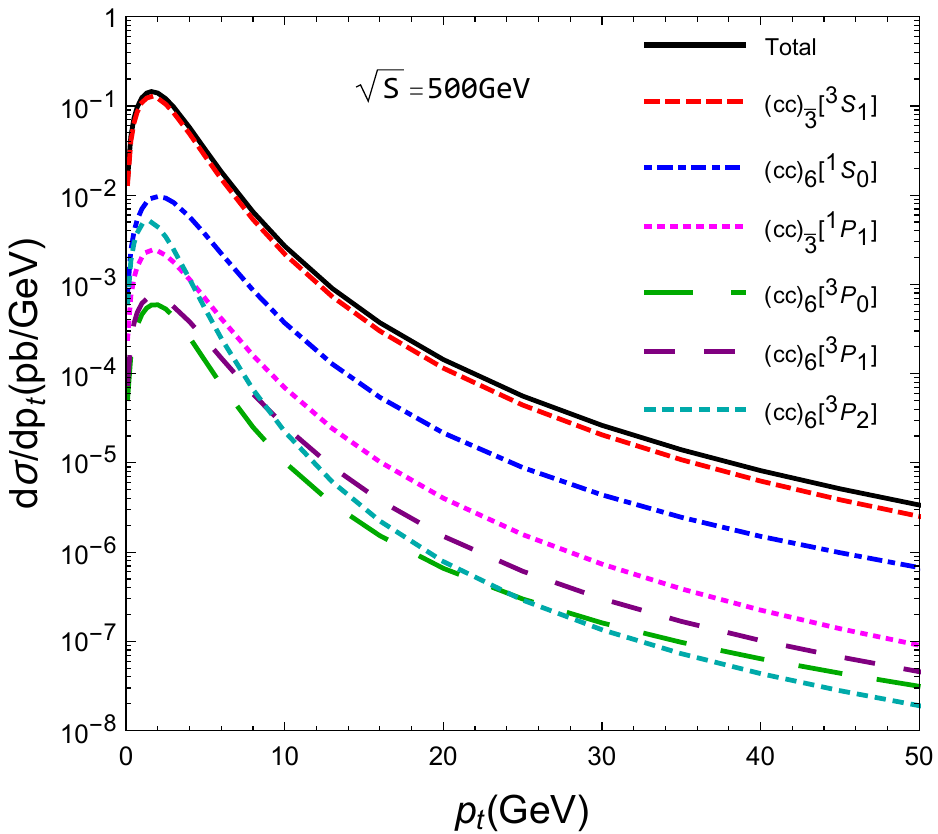}
	\includegraphics[width=.31\textwidth]{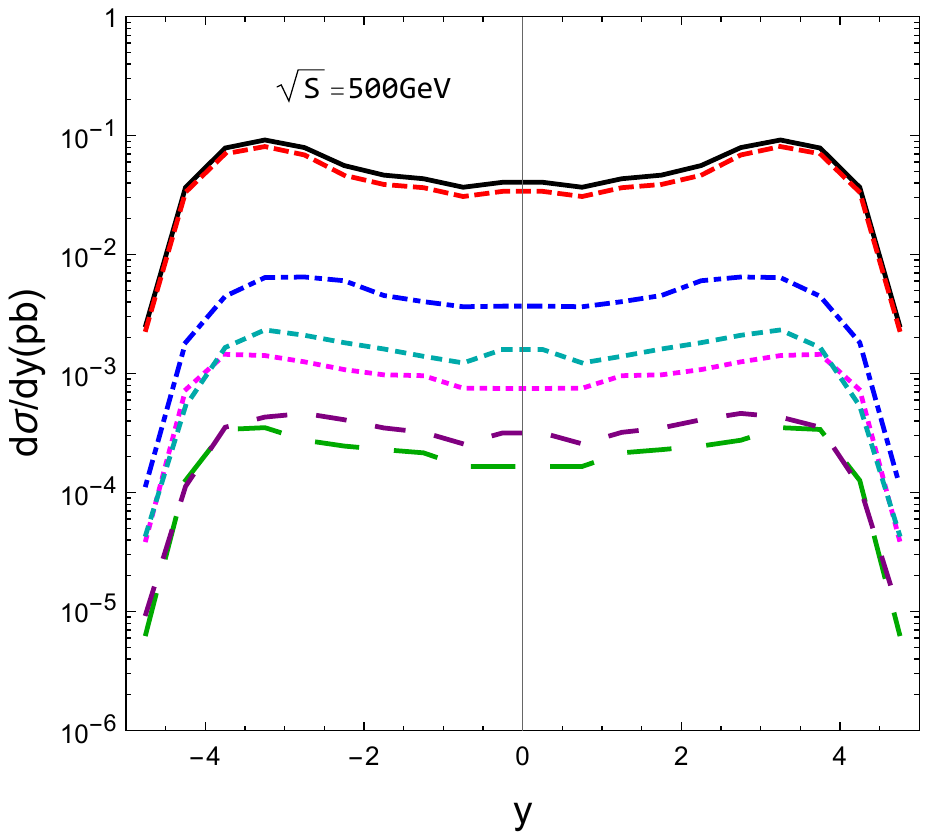}
	\includegraphics[width=.31\textwidth]{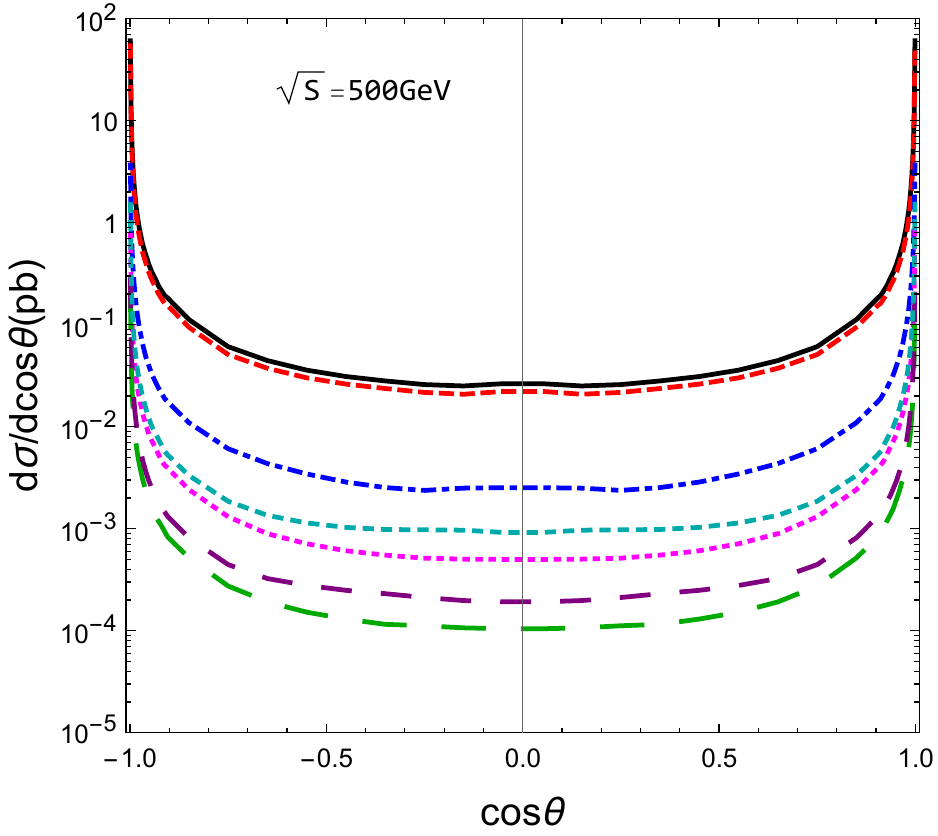}
	\caption{\label{fig:33} Kinematic distributions for the photoproduction of $\Xi_{cc}$ at future $e^+e^-$ collider($\sqrt{S}=$$500$ $\mathrm{GeV}$). Contributions from different diquark states are displayed individually. The $y$ and $\cos\theta$ curves use same legends as those of $p_t$.}
\end{figure}

To reveal more characteristics of $\Xi_{c c}$ photoproduction at the $e^+e^-$ collider, we have computed the differential distributions at $\sqrt{S}=500\,\mathrm{GeV}$, as illustrated in Fig.\ref{fig:32} and Fig.\ref{fig:33}.
Fig.~\ref{fig:32} depicts the transverse momentum($p_t$), the rapidity($y$) and $cos\theta$ distributions, featuring distinct representations of contributions originating from different channels. 
Here, $\theta$ represents the angle between $\Xi_{cc}$ and the $e^+e^-$ beams.
The $\gamma+g$ channels exert their dominance in the lower $p_t$ region, gradually passing the torch to the $\gamma+\gamma$ channels as the $p_t$ values increase.
In practical experiments, there may not be a sufficient number of events in the high $p_t$ region to attain precise measurements. Consequently, it becomes necessary to consider the single resolved channel $\gamma+g$ when performing photoproduction calculations.
In contrast to the curves for the $p_t$ distribution, in the rapidity and angular distributions, the curves for the two production channels do not intersect.
Throughout the entire rapidity distribution range, the contributions of $\gamma+g$ consistently surpass those of $\gamma+\gamma$.
The same pattern is also evident in the angular distribution curves.
Fig.\ref{fig:33} displays the contribution curves for different intermediate diquark states.
In each $p_t$ distribution, a noticeable peak emerges at approximately several GeV, followed by a logarithmic decline in the high $p_t$ region.
In all three distinct kinematic distributions, it is consistently evident that the ${}_{\bar{\textbf{3}}}[{}^3S_1]$ configurations maintain prominence across the entire range, while contributions from other states are small.

\begin{table}
	\centering
	\begin{tabular}{|c|cccccc|}
		\hline
		$m_c(\mathrm{GeV})$ & $(cc)_{\bar{\textbf{3}}}[{}^3S_1]$ & $(cc)_{\textbf{6}}[{}^1S_0]$
		& $(cc)_{\bar{\textbf{3}}}[{}^1P_1]$
		& $(cc)_{\textbf{6}}[{}^3P_0]$ & $(cc)_{\textbf{6}}[{}^3P_1]$ & $(cc)_{\textbf{6}}[{}^3P_2]$\\
		\hline
		1.7 & $609.62$ & $56.85$ & $14.50$ & $3.28$ & $4.67$ & $22.27$\\
		1.8 &  $442.71$ & $41.24$ & $9.42$ & $2.12$ & $3.03$ & $14.32$  \\
		1.9 & $328.38$ & $30.53$ & $6.24$ & $1.40$ & $2.0$ & $9.45$ \\
		\hline
	\end{tabular}
	\caption{\label{tab:uncer-mc}The total cross sections (in unit of fb) under different $m_c$ at $\sqrt{S}=500\mathrm{~GeV}$.}
\end{table}
\begin{table}
	\centering
	\begin{tabular}{|c|cccccc|}
		\hline
		${\cal C}$ & $(cc)_{\bar{\textbf{3}}}[{}^3S_1]$ & $(cc)_{\textbf{6}}[{}^1S_0]$
		& $(cc)_{\bar{\textbf{3}}}[{}^1P_1]$
		& $(cc)_{\textbf{6}}[{}^3P_0]$ & $(cc)_{\textbf{6}}[{}^3P_1]$ & $(cc)_{\textbf{6}}[{}^3P_2]$\\
		\hline
		0.5 & $531.74$ & $50.46$ & $11.44$ & $2.47$ & $3.57$ & $16.40$\\
		1.0 &  $442.71$ & $41.24$ & $9.42$ & $2.12$ & $3.03$ & $14.32$ \\
		2.0 & $376.96$ & $35.12$ & $8.0$ & $1.84$ & $2.59$ & $12.61$ \\
		\hline
	\end{tabular}
	\caption{\label{tab:uncer-mu}The total cross sections (in unit of fb) under various $\mu(={\cal C}\sqrt{M^2_{\Xi_{cc}}+p^2_t}$ with ${\cal C}=0.5,1,2$) at $\sqrt{S}=500\mathrm{~GeV}$.}
\end{table}

Finally, we delve into a brief discussion of the theoretical uncertainties inherent in our calculations, stemming from three main sources: the heavy quark mass $m_c$, the renormalization scale $\mu$, and the LDMEs. It's worth noting that uncertainties originating from $h_{\bar{\textbf{3}}}$ and $h_{\textbf{6}}$ have been omitted due to the lack of reported errors in the literature. As mentioned earlier, these coefficients represent global factors, and their influence on production outcomes can be further refined with more precise values.
Table~\ref{tab:uncer-mc} demonstrates the impact of changing the value of $m_c$ within the range of $1.8\pm0.1\mathrm{~GeV}$ while keeping $\mu=\sqrt{M^2_{\Xi_{cc}}+p^2_t}$ constant.
As observed in the table, even small variations in the heavy quark mass can result in substantial fluctuations in cross-section values. For example, in Table~\ref{tab:uncer-mc}, the cross section for $(cc)_{\bar{\textbf{3}}}[{}^3S_1]$ varies by approximately 46\% with only a 12\% change in $m_c$.
This notable sensitivity becomes evident when analyzing the relevant Feynman diagrams, as exemplified in Fig.~\ref{fig:fd}.
For photoproduction of $\Xi_{cc}$ considered here, it is noted that the final particles involved in the short-distance processes consist exclusively of $c$ and $\bar{c}$, while the internal lines are composed solely of charm and gluon propagators. Therefore, the significant impact of heavy quark masses on the cross section appears to be a reasonable outcome.

Table~\ref{tab:uncer-mu} evaluates the sensitivity to the renormalization scale ($\mu={\cal C}\sqrt{M^2_{\Xi_{cc}}+p^2_t}$, where ${\cal C}=0.5,1,2$), while keeping the value of $m_c$ fixed at $1.8\mathrm{~GeV}$.
Clearly, there is a significant dependence on the renormalization scale, which could suggest the importance of next-to-leading order corrections in $\alpha_s$. As we confront real-world measurements in the future, high-order calculations become imperative.
Taking into account the aforementioned uncertainties, our leading-order calculation results may vary by approximately one order of magnitude. Despite this range of variability, the photoproduction rates of doubly charmed baryons remain significant.

\section{Summary}
\label{sec:4}

In this work, we have investigated the $\Xi_{cc}$ photoproduction within the framework of non-relativistic QCD specifically focusing on future $e^+e^-$ colliders. 
Two dominant photoproduction processes are considered, i.e., $\gamma+\gamma \rightarrow \Xi_{c c} +\bar{c}+\bar{c}$ and $\gamma+g \rightarrow \Xi_{c c} +\bar{c}+\bar{c}$.
Four $P$-wave diquark states are included in the calculation and they are $(cc)_{\bar{\textbf{3}}}[{}^1P_1]$,
$(cc)_{\textbf{6}}[{}^3P_0]$,
$(cc)_{\textbf{6}}[{}^3P_1]$ and
$(cc)_{\textbf{6}}[{}^3P_2]$.
Upon assuming $h_{\textbf{6}}=h_{\bar{\textbf{3}}}$, the results demonstrate the photoproduction of $P$-wave $\Xi_{c c}$ is about one order lower than that of the $S$-wave.
Specifically, at a center-of-mass energy of $\sqrt{S}=500\,\mathrm{GeV}$, the cross section for $P$-wave $\Xi_{c c}$ production is approximately 6\% of that for $S$-wave production.
The numerical results further emphasize the crucial role played by the single resolved photoproduction channel $\gamma+g$ in the overall photoproduction process.
It gains increasing significance as the collision energy rises.
Assuming an integrated luminosity for future $e^+e^-$ collisions on the order of $\mathcal{O}(10^4)\mathrm{~fb^{-1}}$, approximately $4.8\times10^6$ $S$-wave $\Xi_{cc}$ and $2.9\times10^5$ $P$-wave $\Xi_{cc}$ baryons would be expected to be produced at a collision energy of $\sqrt{S}=500\mathrm{~GeV}$.
The excited doubly charmed baryons are likely to decay into the ground state with nearly 100\% probability. Therefore, when faced with precise real-world measurements, their contributions should be thoroughly examined and taken into account.

\acknowledgments

This work was supported in part by the Natural Science Foundation of China under Grants No. 12305083, No. 12147116, No. 12175025, No. 12005028 and No. 12147102, and by the Fundamental Research Funds for the Central Universities under Grant No. 2020CQJQY-Z003.

\bibliographystyle{JHEP.bst}

\begin{thebibliography}{10}
	
	\bibitem{Mattson:2002vu}
	{\scshape SELEX} collaboration, \emph{{First Observation of the Doubly Charmed
			Baryon $\Xi^+_{cc}$}},
	\href{https://doi.org/10.1103/PhysRevLett.89.112001}{\emph{Phys. Rev. Lett.}
		{\bfseries 89} (2002) 112001}
	[\href{https://arxiv.org/abs/hep-ex/0208014}{{\ttfamily hep-ex/0208014}}].
	
	\bibitem{Ocherashvili:2004hi}
	{\scshape SELEX} collaboration, \emph{{Confirmation of the double charm baryon
			Xi+(cc)(3520) via its decay to p D+ K-}},
	\href{https://doi.org/10.1016/j.physletb.2005.09.043}{\emph{Phys. Lett. B}
		{\bfseries 628} (2005) 18}
	[\href{https://arxiv.org/abs/hep-ex/0406033}{{\ttfamily hep-ex/0406033}}].
	
	\bibitem{Aaij:2018gfl}
	{\scshape LHCb} collaboration, \emph{{First Observation of the Doubly Charmed
			Baryon Decay $\Xi_{cc}^{++}\rightarrow \Xi_{c}^{+}\pi^{+}$}},
	\href{https://doi.org/10.1103/PhysRevLett.121.162002}{\emph{Phys. Rev. Lett.}
		{\bfseries 121} (2018) 162002}
	[\href{https://arxiv.org/abs/1807.01919}{{\ttfamily 1807.01919}}].
	
	\bibitem{Aaij:2018wzf}
	{\scshape LHCb} collaboration, \emph{{Measurement of the Lifetime of the Doubly
			Charmed Baryon $\Xi_{cc}^{++}$}},
	\href{https://doi.org/10.1103/PhysRevLett.121.052002}{\emph{Phys. Rev. Lett.}
		{\bfseries 121} (2018) 052002}
	[\href{https://arxiv.org/abs/1806.02744}{{\ttfamily 1806.02744}}].
	
	\bibitem{Ma:2003zk}
	J.P.~Ma and Z.G.~Si, \emph{{Factorization approach for inclusive production of
			doubly heavy baryon}},
	\href{https://doi.org/10.1016/j.physletb.2003.06.064}{\emph{Phys. Lett. B}
		{\bfseries 568} (2003) 135}
	[\href{https://arxiv.org/abs/hep-ph/0305079}{{\ttfamily hep-ph/0305079}}].
	
	\bibitem{Bodwin:1994jh}
	G.T.~Bodwin, E.~Braaten and G.P.~Lepage, \emph{{Rigorous QCD analysis of
			inclusive annihilation and production of heavy quarkonium}},
	\href{https://doi.org/10.1103/PhysRevD.55.5853,
		10.1103/PhysRevD.51.1125}{\emph{Phys. Rev.} {\bfseries D51} (1995) 1125}
	[\href{https://arxiv.org/abs/hep-ph/9407339}{{\ttfamily hep-ph/9407339}}].
	
	\bibitem{Baranov:1995rc}
	S.P.~Baranov, \emph{{On the production of doubly flavored baryons in p p, e p
			and gamma gamma collisions}},
	\href{https://doi.org/10.1103/PhysRevD.54.3228}{\emph{Phys. Rev. D}
		{\bfseries 54} (1996) 3228}.
	
	\bibitem{Berezhnoy:1996an}
	A.V.~Berezhnoy, V.V.~Kiselev and A.K.~Likhoded, \emph{{Photonic production of
			S- and P wave B/c states and doubly heavy baryons}},
	\href{https://doi.org/10.1007/s002180050152}{\emph{Z. Phys. A} {\bfseries
			356} (1996) 89}.
	
	\bibitem{Jiang:2012jt}
	J.~Jiang, X.-G.~Wu, Q.-L.~Liao, X.-C.~Zheng and Z.-Y.~Fang, \emph{{Doubly Heavy
			Baryon Production at A High Luminosity $e^+ e^-$ Collider}},
	\href{https://doi.org/10.1103/PhysRevD.86.054021}{\emph{Phys. Rev. D}
		{\bfseries 86} (2012) 054021}
	[\href{https://arxiv.org/abs/1208.3051}{{\ttfamily 1208.3051}}].
	
	\bibitem{Jiang:2013ej}
	J.~Jiang, X.-G.~Wu, S.-M.~Wang, J.-W.~Zhang and Z.-Y.~Fang, \emph{{A Further
			Study on the Doubly Heavy Baryon Production around the $Z^0$ Peak at A High
			Luminosity $e^+ e^-$ Collider}},
	\href{https://doi.org/10.1103/PhysRevD.87.054027}{\emph{Phys. Rev. D}
		{\bfseries 87} (2013) 054027}
	[\href{https://arxiv.org/abs/1302.0601}{{\ttfamily 1302.0601}}].
	
	\bibitem{Chen:2014frw}
	G.~Chen, X.-G.~Wu, Z.~Sun, Y.~Ma and H.-B.~Fu, \emph{{Photoproduction of doubly
			heavy baryon at the ILC}},
	\href{https://doi.org/10.1007/JHEP12(2014)018}{\emph{JHEP} {\bfseries 12}
		(2014) 018} [\href{https://arxiv.org/abs/1408.4615}{{\ttfamily 1408.4615}}].
	
	\bibitem{Yang:2014ita}
	Z.-J.~Yang, P.-F.~Zhang and Y.-J.~Zheng, \emph{{Doubly Heavy Baryon Production
			in $e^{+}e^{-}$ Annihilation}},
	\href{https://doi.org/10.1088/0256-307X/31/5/051301}{\emph{Chin. Phys. Lett.}
		{\bfseries 31} (2014) 051301}.
	
	\bibitem{Yang:2014tca}
	Z.-J.~Yang and X.-X.~Zhao, \emph{{The Production of $\Xi_{bb}$ at Photon
			Collider}}, \href{https://doi.org/10.1088/0256-307X/31/9/091301}{\emph{Chin.
			Phys. Lett.} {\bfseries 31} (2014) 091301}
	[\href{https://arxiv.org/abs/1408.5584}{{\ttfamily 1408.5584}}].
	
	\bibitem{Zheng:2015ixa}
	X.-C.~Zheng, C.-H.~Chang and Z.~Pan, \emph{{Production of doubly heavy-flavored
			hadrons at $e^+e^-$ colliders}},
	\href{https://doi.org/10.1103/PhysRevD.93.034019}{\emph{Phys. Rev. D}
		{\bfseries 93} (2016) 034019}
	[\href{https://arxiv.org/abs/1510.06808}{{\ttfamily 1510.06808}}].
	
	\bibitem{Bi:2017nzv}
	H.-Y.~Bi, R.-Y.~Zhang, X.-G.~Wu, W.-G.~Ma, X.-Z.~Li and S.~Owusu,
	\emph{{Photoproduction of doubly heavy baryon at the LHeC}},
	\href{https://doi.org/10.1103/PhysRevD.95.074020}{\emph{Phys. Rev. D}
		{\bfseries 95} (2017) 074020}
	[\href{https://arxiv.org/abs/1702.07181}{{\ttfamily 1702.07181}}].
	
	\bibitem{Sun:2020mvl}
	Z.~Sun and X.-G.~Wu, \emph{{The production of the doubly charmed baryon in
			deeply inelastic $ep$ scattering at the Large Hadron Electron Collider}},
	\href{https://doi.org/10.1007/JHEP07(2020)034}{\emph{JHEP} {\bfseries 07}
		(2020) 034} [\href{https://arxiv.org/abs/2004.01012}{{\ttfamily
			2004.01012}}].
	
	\bibitem{Chen:2014hqa}
	G.~Chen, X.-G.~Wu, J.-W.~Zhang, H.-Y.~Han and H.-B.~Fu, \emph{{Hadronic
			production of $\Xi_{cc}$ at a fixed-target experiment at the LHC}},
	\href{https://doi.org/10.1103/PhysRevD.89.074020}{\emph{Phys. Rev. D}
		{\bfseries 89} (2014) 074020}
	[\href{https://arxiv.org/abs/1401.6269}{{\ttfamily 1401.6269}}].
	
	\bibitem{Chen:2019ykv}
	G.~Chen, X.-G.~Wu and S.~Xu, \emph{{Impacts of the intrinsic charm content of
			the proton on the $\Xi_{cc}$ hadroproduction at a fixed target experiment at
			the LHC}}, \href{https://doi.org/10.1103/PhysRevD.100.054022}{\emph{Phys.
			Rev. D} {\bfseries 100} (2019) 054022}
	[\href{https://arxiv.org/abs/1903.00722}{{\ttfamily 1903.00722}}].
	
	\bibitem{Chen:2018koh}
	G.~Chen, C.-H.~Chang and X.-G.~Wu, \emph{{Hadronic production of the doubly
			charmed baryon via the proton\textendash{}nucleus and the
			nucleus\textendash{}nucleus collisions at the RHIC and LHC}},
	\href{https://doi.org/10.1140/epjc/s10052-018-6283-1}{\emph{Eur. Phys. J. C}
		{\bfseries 78} (2018) 801}
	[\href{https://arxiv.org/abs/1808.03174}{{\ttfamily 1808.03174}}].
	
	\bibitem{Martynenko:2014ola}
	A.P.~Martynenko and A.M.~Trunin, \emph{{Pair double heavy diquark production in
			high energy proton\textendash{}proton collisions}},
	\href{https://doi.org/10.1140/epjc/s10052-015-3358-0}{\emph{Eur. Phys. J. C}
		{\bfseries 75} (2015) 138} [\href{https://arxiv.org/abs/1405.0969}{{\ttfamily
			1405.0969}}].
	
	\bibitem{Koshkarev:2016acq}
	S.~Koshkarev, \emph{{Production of the Doubly Heavy Baryons, $B_c$ Meson and
			the All-charm Tetraquark at AFTER@LHC with Double Intrinsic Heavy
			Mechanism}}, \href{https://doi.org/10.5506/APhysPolB.48.163}{\emph{Acta Phys.
			Polon. B} {\bfseries 48} (2017) 163}
	[\href{https://arxiv.org/abs/1610.06125}{{\ttfamily 1610.06125}}].
	
	\bibitem{Koshkarev:2016rci}
	S.~Koshkarev and V.~Anikeev, \emph{{Production of the doubly charmed baryons at
			the SELEX experiment \textendash{} The double intrinsic charm approach}},
	\href{https://doi.org/10.1016/j.physletb.2016.12.010}{\emph{Phys. Lett. B}
		{\bfseries 765} (2017) 171}
	[\href{https://arxiv.org/abs/1605.03070}{{\ttfamily 1605.03070}}].
	
	\bibitem{Groote:2017szb}
	S.~Groote and S.~Koshkarev, \emph{{Production of doubly charmed baryons nearly
			at rest}}, \href{https://doi.org/10.1140/epjc/s10052-017-5086-0}{\emph{Eur.
			Phys. J. C} {\bfseries 77} (2017) 509}
	[\href{https://arxiv.org/abs/1704.02850}{{\ttfamily 1704.02850}}].
	
	\bibitem{Berezhnoy:2018bde}
	A.V.~Berezhnoy, A.K.~Likhoded and A.V.~Luchinsky, \emph{{Doubly heavy baryons
			at the LHC}}, \href{https://doi.org/10.1103/PhysRevD.98.113004}{\emph{Phys.
			Rev. D} {\bfseries 98} (2018) 113004}
	[\href{https://arxiv.org/abs/1809.10058}{{\ttfamily 1809.10058}}].
	
	\bibitem{Brodsky:2017ntu}
	S.J.~Brodsky, S.~Groote and S.~Koshkarev, \emph{{Resolving the
			SELEX\textendash{}LHCb double-charm baryon conflict: the impact of intrinsic
			heavy-quark hadroproduction and supersymmetric light-front holographic QCD}},
	\href{https://doi.org/10.1140/epjc/s10052-018-5955-1}{\emph{Eur. Phys. J. C}
		{\bfseries 78} (2018) 483}
	[\href{https://arxiv.org/abs/1709.09903}{{\ttfamily 1709.09903}}].
	
	\bibitem{Berezhnoy:2018krl}
	A.V.~Berezhnoy, I.N.~Belov and A.K.~Likhoded, \emph{{Production of doubly
			charmed baryons with the excited heavy diquark at LHC}},
	\href{https://doi.org/10.1142/S0217751X19500386}{\emph{Int. J. Mod. Phys. A}
		{\bfseries 34} (2019) 1950038}
	[\href{https://arxiv.org/abs/1811.07382}{{\ttfamily 1811.07382}}].
	
	\bibitem{Wu:2019gta}
	X.-G.~Wu, \emph{{A new search for the doubly charmed baryon $\Xi_{cc}^+$ at the
			LHC}}, \href{https://doi.org/10.1007/s11433-019-1478-x}{\emph{Sci. China
			Phys. Mech. Astron.} {\bfseries 63} (2020) 221063}
	[\href{https://arxiv.org/abs/1912.01953}{{\ttfamily 1912.01953}}].
	
	\bibitem{Qin:2020zlg}
	Q.~Qin, Y.-F.~Shen and F.-S.~Yu, \emph{{Discovery potentials of double-charm
			tetraquarks}}, \href{https://doi.org/10.1088/1674-1137/ac1b97}{\emph{Chin.
			Phys. C} {\bfseries 45} (2021) 103106}
	[\href{https://arxiv.org/abs/2008.08026}{{\ttfamily 2008.08026}}].
	
	\bibitem{Niu:2018ycb}
	J.-J.~Niu, L.~Guo, H.-H.~Ma, X.-G.~Wu and X.-C.~Zheng, \emph{{Production of
			semi-inclusive doubly heavy baryons via top-quark decays}},
	\href{https://doi.org/10.1103/PhysRevD.98.094021}{\emph{Phys. Rev. D}
		{\bfseries 98} (2018) 094021}
	[\href{https://arxiv.org/abs/1810.03834}{{\ttfamily 1810.03834}}].
	
	\bibitem{Niu:2019xuq}
	J.-J.~Niu, L.~Guo, H.-H.~Ma and X.-G.~Wu, \emph{{Production of doubly heavy
			baryons via Higgs boson decays}},
	\href{https://doi.org/10.1140/epjc/s10052-019-6842-0}{\emph{Eur. Phys. J. C}
		{\bfseries 79} (2019) 339}
	[\href{https://arxiv.org/abs/1904.02339}{{\ttfamily 1904.02339}}].
	
	\bibitem{Zhang:2022jst}
	P.-H.~Zhang, L.~Guo, X.-C.~Zheng and Q.-W.~Ke, \emph{{Excited doubly heavy
			baryon production via $W^+$ boson decays}},
	\href{https://doi.org/10.1103/PhysRevD.105.034016}{\emph{Phys. Rev. D}
		{\bfseries 105} (2022) 034016}
	[\href{https://arxiv.org/abs/2202.01579}{{\ttfamily 2202.01579}}].
	
	\bibitem{Luo:2022jxq}
	X.~Luo, Y.-Z.~Jiang, G.-Y.~Zhang and Z.~Sun, \emph{{Doubly-charmed baryon
			production in $Z$ boson decay}},
	\href{https://arxiv.org/abs/2206.05965}{{\ttfamily 2206.05965}}.
	
	\bibitem{Luo:2022lcj}
	X.~Luo, H.-B.~Fu and H.-J.~Tian, \emph{{Investigation of Z-boson decay into and
			baryons within the NRQCD factorization approach*}},
	\href{https://doi.org/10.1088/1674-1137/acbc0e}{\emph{Chin. Phys. C}
		{\bfseries 47} (2023) 053102}
	[\href{https://arxiv.org/abs/2208.07520}{{\ttfamily 2208.07520}}].
	
	\bibitem{Ma:2022cgt}
	H.-H.~Ma, J.-J.~Niu and X.-C.~Zheng, \emph{{Excited doubly heavy baryons
			production via top-quark decays}},
	\href{https://doi.org/10.1103/PhysRevD.107.014006}{\emph{Phys. Rev. D}
		{\bfseries 107} (2023) 014006}
	[\href{https://arxiv.org/abs/2210.03306}{{\ttfamily 2210.03306}}].
	
	\bibitem{Chang:2007pp}
	C.-H.~Chang, J.-X.~Wang and X.-G.~Wu, \emph{{GENXICC: A Generator for hadronic
			production of the double heavy baryons Xi(cc), Xi(bc) and Xi(bb)}},
	\href{https://doi.org/10.1016/j.cpc.2007.05.012}{\emph{Comput. Phys. Commun.}
		{\bfseries 177} (2007) 467}
	[\href{https://arxiv.org/abs/hep-ph/0702054}{{\ttfamily hep-ph/0702054}}].
	
	\bibitem{Chang:2009va}
	C.-H.~Chang, J.-X.~Wang and X.-G.~Wu, \emph{{GENXICC2.0: An Upgraded Version of
			the Generator for Hadronic Production of Double Heavy Baryons Xi(cc), Xi(bc)
			and Xi(bb)}}, \href{https://doi.org/10.1016/j.cpc.2010.02.008}{\emph{Comput.
			Phys. Commun.} {\bfseries 181} (2010) 1144}
	[\href{https://arxiv.org/abs/0910.4462}{{\ttfamily 0910.4462}}].
	
	\bibitem{Wang:2012vj}
	X.-Y.~Wang and X.-G.~Wu, \emph{{GENXICC2.1: An Improved Version of GENXICC for
			Hadronic Production of Doubly Heavy Baryons}},
	\href{https://doi.org/10.1016/j.cpc.2012.10.022}{\emph{Comput. Phys. Commun.}
		{\bfseries 184} (2013) 1070}
	[\href{https://arxiv.org/abs/1210.3458}{{\ttfamily 1210.3458}}].
	
	\bibitem{FCC:2018evy}
	{\scshape FCC} collaboration, \emph{{FCC-ee: The Lepton Collider}: {Future
			Circular Collider Conceptual Design Report Volume 2}},
	\href{https://doi.org/10.1140/epjst/e2019-900045-4}{\emph{Eur. Phys. J. ST}
		{\bfseries 228} (2019) 261}.
	
	\bibitem{CEPCStudyGroup:2018rmc}
	C.S.~Group, \emph{{CEPC Conceptual Design Report: Volume 1 - Accelerator}},
	\href{https://arxiv.org/abs/1809.00285}{{\ttfamily 1809.00285}}.
	
	\bibitem{CEPCStudyGroup:2018ghi}
	{\scshape CEPC Study Group} collaboration, \emph{{CEPC Conceptual Design
			Report: Volume 2 - Physics \& Detector}},
	\href{https://arxiv.org/abs/1811.10545}{{\ttfamily 1811.10545}}.
	
	\bibitem{ILC:2007bjz}
	{\scshape ILC} collaboration, \emph{{International Linear Collider Reference
			Design Report Volume 2: Physics at the ILC}},
	\href{https://arxiv.org/abs/0709.1893}{{\ttfamily 0709.1893}}.
	
	\bibitem{Erler:2000jg}
	J.~Erler, S.~Heinemeyer, W.~Hollik, G.~Weiglein and P.M.~Zerwas, \emph{{Physics
			impact of GigaZ}},
	\href{https://doi.org/10.1016/S0370-2693(00)00749-8}{\emph{Phys. Lett. B}
		{\bfseries 486} (2000) 125}
	[\href{https://arxiv.org/abs/hep-ph/0005024}{{\ttfamily hep-ph/0005024}}].
	
	\bibitem{Klasen:2001cu}
	M.~Klasen, B.A.~Kniehl, L.N.~Mihaila and M.~Steinhauser, \emph{{Evidence for
			color octet mechanism from CERN LEP-2 $\gamma \gamma \to J/\psi$ + $X$
			data}}, \href{https://doi.org/10.1103/PhysRevLett.89.032001}{\emph{Phys. Rev.
			Lett.} {\bfseries 89} (2002) 032001}
	[\href{https://arxiv.org/abs/hep-ph/0112259}{{\ttfamily hep-ph/0112259}}].
	
	\bibitem{Li:2009zzu}
	R.~Li and K.-T.~Chao, \emph{{Photoproduction of $J/psi$ in association with a
			$c \bar{c}$ pair}},
	\href{https://doi.org/10.1103/PhysRevD.79.114020}{\emph{Phys. Rev. D}
		{\bfseries 79} (2009) 114020}
	[\href{https://arxiv.org/abs/0904.1643}{{\ttfamily 0904.1643}}].
	
	\bibitem{Zhan:2020ugq}
	X.-J.~Zhan and J.-X.~Wang, \emph{{Prompt $J/\psi$ photoproduction within the
			non-relativistic QCD framework at the CEPC}},
	\href{https://doi.org/10.1140/epjc/s10052-020-8276-0}{\emph{Eur. Phys. J. C}
		{\bfseries 80} (2020) 740}
	[\href{https://arxiv.org/abs/2005.08816}{{\ttfamily 2005.08816}}].
	
	\bibitem{Zhan:2021dlu}
	X.-J.~Zhan and J.-X.~Wang, \emph{{Inclusive $\Upsilon (1S,2S,3S)$
			photoproduction at the CEPC}},
	\href{https://doi.org/10.1088/1674-1137/abce11}{\emph{Chin. Phys. C}
		{\bfseries 45} (2021) 023112}.
	
	\bibitem{Zhan:2022nck}
	X.-J.~Zhan, X.-G.~Wu and X.-C.~Zheng, \emph{{Inclusive J/\ensuremath{\psi}
			photoproduction at the ILC within the framework of non-relativistic QCD}},
	\href{https://doi.org/10.1007/JHEP09(2022)050}{\emph{JHEP} {\bfseries 09}
		(2022) 050} [\href{https://arxiv.org/abs/2207.01763}{{\ttfamily
			2207.01763}}].
	
	\bibitem{Zhan:2022etq}
	X.-J.~Zhan, X.-G.~Wu and X.-C.~Zheng, \emph{{Photoproduction of the Bc meson at
			future e+e- colliders}},
	\href{https://doi.org/10.1103/PhysRevD.106.094036}{\emph{Phys. Rev. D}
		{\bfseries 106} (2022) 094036}
	[\href{https://arxiv.org/abs/2211.09003}{{\ttfamily 2211.09003}}].
	
	\bibitem{Ginzburg:1981vm}
	I.F.~Ginzburg, G.L.~Kotkin, V.G.~Serbo and V.I.~Telnov, \emph{{Colliding gamma
			e and gamma gamma Beams Based on the Single Pass Accelerators (of Vlepp
			Type)}}, \href{https://doi.org/10.1016/0167-5087(83)90173-4}{\emph{Nucl.
			Instrum. Meth.} {\bfseries 205} (1983) 47}.
	
	\bibitem{Telnov:1989sd}
	V.I.~Telnov, \emph{{Problems of Obtaining $\gamma \gamma$ and $\gamma \epsilon$
			Colliding Beams at Linear Colliders}},
	\href{https://doi.org/10.1016/0168-9002(90)91826-W}{\emph{Nucl. Instrum.
			Meth. A} {\bfseries 294} (1990) 72}.
	
	\bibitem{Gluck:1999ub}
	M.~Gluck, E.~Reya and I.~Schienbein, \emph{{Radiatively generated parton
			distributions of real and virtual photons}},
	\href{https://doi.org/10.1103/PhysRevD.60.054019,
		10.1103/PhysRevD.62.019902}{\emph{Phys. Rev.} {\bfseries D60} (1999) 054019}
	[\href{https://arxiv.org/abs/hep-ph/9903337}{{\ttfamily hep-ph/9903337}}].
	
	\bibitem{Falk:1993gb}
	A.F.~Falk, M.E.~Luke, M.J.~Savage and M.B.~Wise, \emph{{Heavy quark
			fragmentation to baryons containing two heavy quarks}},
	\href{https://doi.org/10.1103/PhysRevD.49.555}{\emph{Phys. Rev. D} {\bfseries
			49} (1994) 555} [\href{https://arxiv.org/abs/hep-ph/9305315}{{\ttfamily
			hep-ph/9305315}}].
	
	\bibitem{Kiselev:1994pu}
	V.V.~Kiselev, A.K.~Likhoded and M.V.~Shevlyagin, \emph{{Double charmed baryon
			production at B factory}},
	\href{https://doi.org/10.1016/0370-2693(94)91273-4}{\emph{Phys. Lett. B}
		{\bfseries 332} (1994) 411}
	[\href{https://arxiv.org/abs/hep-ph/9408407}{{\ttfamily hep-ph/9408407}}].
	
	\bibitem{Bagan:1994dy}
	E.~Bagan, H.G.~Dosch, P.~Gosdzinsky, S.~Narison and J.M.~Richard,
	\emph{{Hadrons with charm and beauty}},
	\href{https://doi.org/10.1007/BF01557235}{\emph{Z. Phys. C} {\bfseries 64}
		(1994) 57} [\href{https://arxiv.org/abs/hep-ph/9403208}{{\ttfamily
			hep-ph/9403208}}].
	
	\bibitem{Berezhnoy:1998aa}
	A.~Berezhnoy, V.~Kiselev, A.~Likhoded and A.~Onishchenko, \emph{{Doubly charmed
			baryon production in hadronic experiments}},
	\href{https://doi.org/10.1103/PhysRevD.57.4385}{\emph{Phys. Rev. D}
		{\bfseries 57} (1998) 4385}
	[\href{https://arxiv.org/abs/hep-ph/9710339}{{\ttfamily hep-ph/9710339}}].
	
	\bibitem{Binosi:2003yf}
	D.~Binosi and L.~Theussl, \emph{{JaxoDraw: A Graphical user interface for
			drawing Feynman diagrams}},
	\href{https://doi.org/10.1016/j.cpc.2004.05.001}{\emph{Comput. Phys. Commun.}
		{\bfseries 161} (2004) 76}
	[\href{https://arxiv.org/abs/hep-ph/0309015}{{\ttfamily hep-ph/0309015}}].
	
	\bibitem{Kiselev:2002iy}
	V.V.~Kiselev, A.K.~Likhoded, O.N.~Pakhomova and V.A.~Saleev, \emph{{Mass
			spectra of doubly heavy Omega $Q Q^\prime$ baryons}},
	\href{https://doi.org/10.1103/PhysRevD.66.034030}{\emph{Phys. Rev. D}
		{\bfseries 66} (2002) 034030}
	[\href{https://arxiv.org/abs/hep-ph/0206140}{{\ttfamily hep-ph/0206140}}].
	
\end{thebibliography}

\providecommand{\href}[2]{#2}\begingroup\raggedright\endgroup

\end{document}